\documentclass[%
 reprint,
 amsmath,amssymb,
 aps,
prb,
 superscriptaddress
]{revtex4-2}

\usepackage{graphicx}
\usepackage{dcolumn}
\usepackage{bm}
\usepackage{color}
\usepackage{mathrsfs}

\renewcommand{\Re}{\operatorname{Re}}
\renewcommand{\Im}{\operatorname{Im}}

\begin{document}

\preprint{APS/123-QED}

\title{Non-Hermitian waves in a continuous periodic model and application to photonic crystals}

\author{Kazuki Yokomizo}
\affiliation{Condensed Matter Theory Laboratory, RIKEN, 2-1 Hirosawa, Wako, Saitama, 351-0198, Japan}
\author{Taiki Yoda}
\affiliation{NTT Basic Research Laboratories, NTT Corporation, 3-1 Morinosato-Wakamiya, Atsugi-shi, Kanagawa 243-0198, Japan}%
\author{Shuichi Murakami}%
\affiliation{Department of Physics, Tokyo Institute of Technology, 2-12-1 Ookayama, Meguro-ku, Tokyo, 152-8551, Japan}
\affiliation{TIES, Tokyo Institute of Technology, 2-12-1 Ookayama, Meguro-ku, Tokyo, 152-8551, Japan}%




%
\begin{abstract}
In some non-Hermitian systems, the eigenstates in the bulk are localized at the boundaries of the systems. This is called the non-Hermitian skin effect, and it has been studied mostly in discrete systems. In the present work, we study the non-Hermitian skin effect in a continuous periodic model. In a one-dimensional system, we show that the localization lengths are equal for all the eigenstates. Moreover, the localization length and the eigenspectra in a large system are independent of the types of open boundary conditions. These properties are also found in a non-Hermitian photonic crystal. Such remarkable behaviors in a continuous periodic model can be explained in terms of the non-Bloch band theory. By constructing the generalized Brillouin zone for a complex Bloch wave number, we derive the localization length and the eigenspectra under an open boundary condition. Furthermore we show that the generalized Brillouin zone also has various physical properties, such as bulk-edge correspondence.
\end{abstract}
\pacs{Valid PACS appear here}
\maketitle
%
%

\section{\label{sec1}Introduction}
Classical waves, such as elastic waves, acoustic waves and electromagnetic waves, are fundamental research topics in various fields of physics. In periodic structures, these waves form band structures described in terms of the Bloch band theory, and they are controllable. For example, a multilayer of two dielectric media has a complete band gap of electromagnetic waves propagating along the stacking direction~\cite{Joannopoulos2008}. As a result, electromagnetic waves with frequencies within the gap are forbidden to propagate in the photonic crystal.

In recent years, optical systems which incorporate non-Hermiticity, such as gain and loss, have been attracting much attention in theory and in experiment. Interestingly, non-Hermiticity leads to remarkable phenomena which have no counterpart to Hermitian systems. Non-Hermitian systems with $\mathscr{PT}$ symmetry~\cite{Bender1998} are intensively studied in optics, and they result in unidirectional transmission, retroreflection, and so on~\cite{Guo2009,Feng2013,Eichelkraut2013,Xu2016,Wang2019}. In addition, topological edge modes can be created in optical systems with periodically aligned gain and loss, and they can be modulated by gain and loss~\cite{Takata2018,Luo2019,Zeuner2015,Weimann2017,Pan2018,Liu2020,Jeon2020,Kim2020,Gao2021}. Recently, non-Hermitian topology stemming from complex eigenvalues was also investigated~\cite{Zhou2018,KWang2021}. We stress that these results are unique to non-Hermitian systems.
 
In theoretical and experimental research on non-Hermitian systems, a non-Hermitian skin effect plays a crucial role. In non-Hermitian crystals, this effect leads to localization of bulk eigenstates at boundaries of the systems with open boundary conditions~\cite{Yao2018,Yao2018v2,Okuma2020,Yoshida2020,Zhang2020,Kawabata2020,Okugawa2020}. Accordingly, the non-Hermitian skin effect has rich physics, and phenomena caused by this effect are unique to non-Hermitian systems~\cite{Yao2019,Yokomizo2020v2,Li2020,Yokomizo2021v2}. In fact, the non-Hermitian skin effect was experimentally realized in various physical systems~\cite{Brandenbourger2019,Gou2020,Xiao2020,Helbig2020,Hofmann2020,Ghatak2020,Palacios2021,Zhang2021,Chen2021,LZhang2021,Zhou2021}. In particular, an optical system is a good platform to demonstrate phenomena associated with the non-Hermitian skin effect~\cite{Zhu2020,Weidemann2020,Song2020,Wang2021}. Nevertheless, so far, theoretical studies on the non-Hermitian skin effect have been limited mostly to a tight-binding model. Therefore, in a continuous model, such as a photonic crystal, the behavior of bulk eigenstates in the non-Hermitian skin effect is still unclear.

In this paper, we study a non-Hermitian wave in a continuous periodic model. As examples, we focus on a toy model and a photonic crystal and demonstrate that these systems exhibit the non-Hermitian skin effect. Then we find remarkable behavior of the non-Hermitian skin effect. Namely the bands under an open boundary condition are independent of the type of open boundary conditions when the system size is large. Moreover, the localization length is common for all the eigenstates. We explain the behavior of the non-Hermitian skin effect in terms of the non-Bloch band theory proposed in our previous works. Importantly, we calculate the Brillouin zone unique to non-Hermitian systems, called the generalized Brillouin zone, which reproduces the eigenspectra of the system. In this case, the generalized Brillouin zone becomes a circle, which accounts for the constant localization length of the skin modes. Additionally, in the photonic crystal, we establish bulk-edge correspondence between the Zak phase~\cite{Zak1989} defined from the generalized Brillouin zone and the appearance of topological edge modes.

%
%

\section{\label{sec2}Non-Hermitian skin effect}
\begin{figure}[ptb]
\includegraphics[clip,width=0.45\textwidth]{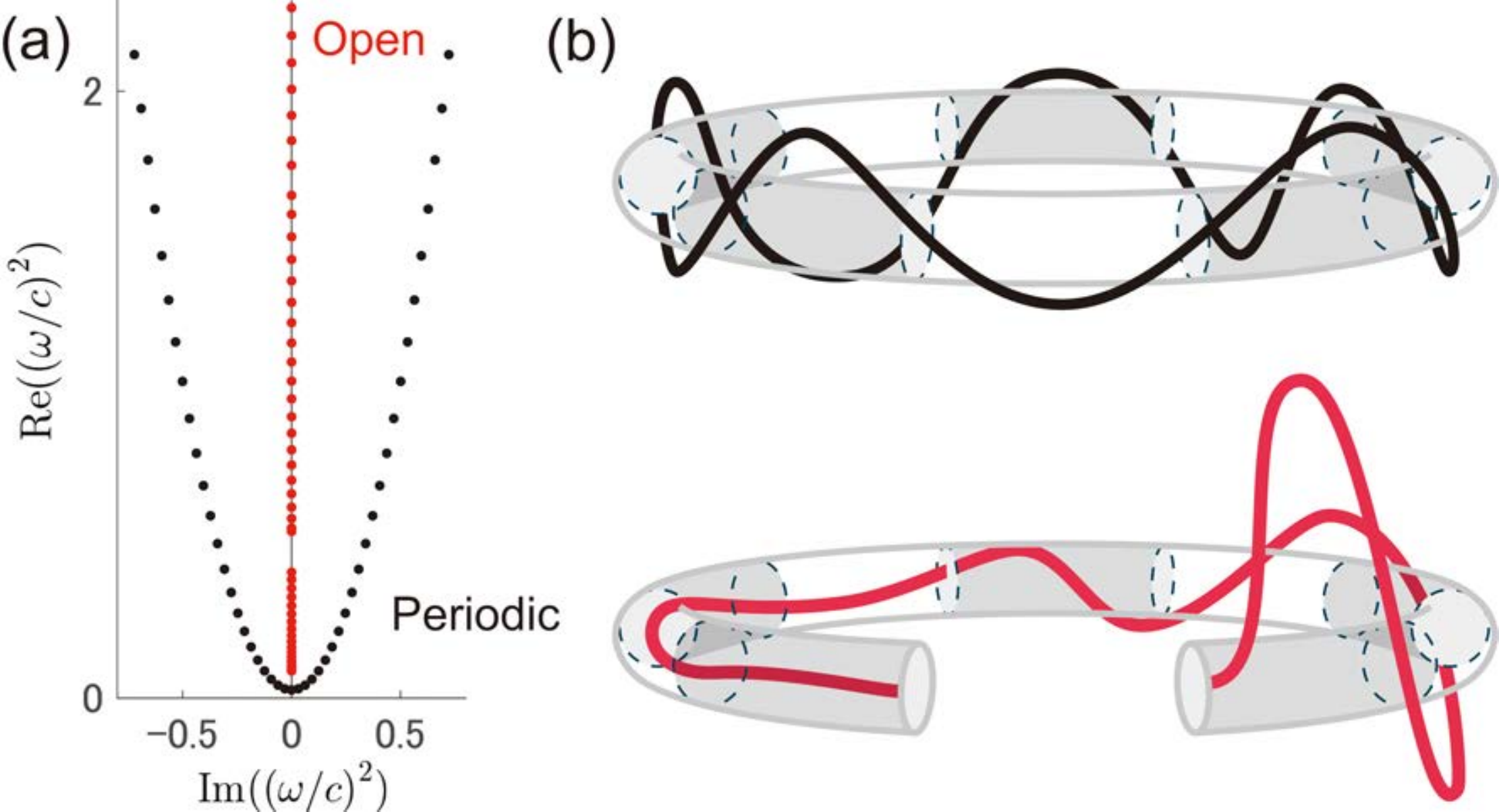}
\caption{\label{fig1}
(a) Eigenvalues with the open boundary condition, $\psi\left(0\right)=\psi\left(L\right)=0$, (red) and those with the periodic boundary condition, $\psi\left(0\right)=\psi\left(L\right)$, (black) in the toy model (\ref{eq3}), where $L$ is the system size. The system parameters are set to be $p=10^{-2},\lambda=10^{-1}$, and the system size is $L=10a$.
(b) Schematic figures of extended eigenstates under a periodic boundary condition (top panel) and localized eigenstates under an open boundary condition (bottom panel) in the non-Hermitian skin effect.}
\end{figure}
\begin{figure}[ptb]
\includegraphics[clip,width=0.45\textwidth]{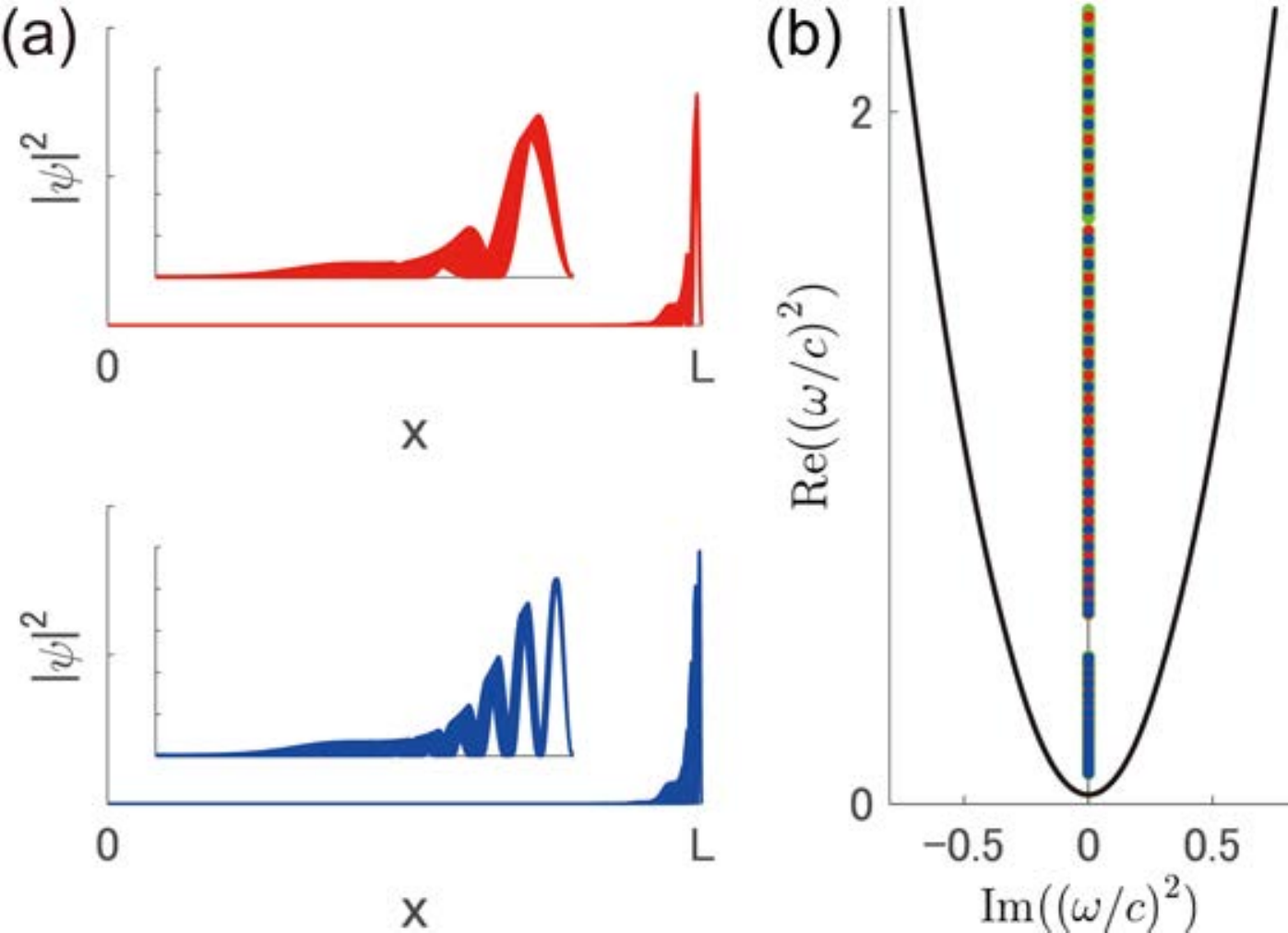}
\caption{\label{fig2}
(a) Spatial distribution of the eigenstates under the Dirichlet boundary condition (red) and under the Neumann boundary condition (blue) in the toy model (\ref{eq3}). These eigenstates are included in the first and second bands. The system parameters are set to be $p=10^{-2},\lambda=10^{-1}$, and the system size is $L=10a$. The insets are extended figures in $x\in\left[L-1,L\right]$.
(b) Eigenvalues with the Dirichlet boundary condition (red dots) and with the Neumann boundary condition (blue dots). For comparison, we also show our results for eigenspectra from the complex Bloch wave number (green) and from the real Bloch wave number (black), corresponding to open and periodic boundary conditions, respectively.}
\end{figure}
Various waves in Hermitian continuous systems in physics are described in terms of the Strum-Liouville equation, which is a standard equation in Hermitian boundary-value problems. In our work, we extend the Strum-Liouville equation to describe non-Hermitian waves. Thus, the wave equation with frequency $\omega$ is given by
\begin{eqnarray}
\left[-\frac{d}{dx}p\left(x\right)\frac{d}{dx}-\frac{i}{2}\left(\lambda_1\left(x\right)\frac{d}{dx}+\frac{d}{dx}\lambda_2\left(x\right)\right)+v\left(x\right)\right]\psi\left(x\right) \nonumber\\
=\left(\frac{\omega}{c}\right)^2\psi\left(x\right), \nonumber\\
\label{eq1}
\end{eqnarray}
where $c$ is a positive constant, and $\psi\left(x\right)$ is the wave function. Here we take the convention of the time dependence of the waves to be $e^{-i\omega t}$. In Eq.~(\ref{eq1}), $p\left(x\right),\lambda_{1,2}\left(x\right)$, and $v\left(x\right)$ are complex periodic functions
\begin{equation}
p\left(x+a\right)=p\left(x\right),\lambda_{1,2}\left(x+a\right)=\lambda_{1,2}\left(x\right),v\left(x+a\right)=v\left(x\right),
\label{eq2}
\end{equation}
where $a$ is a lattice constant. Then the terms including $\lambda_{1,2}\left(x\right)$ express an imaginary gauge potential~\cite{Hatano1996}. We note that the system becomes Hermitian when $p\left(x\right)$ and $v\left(x\right)$ are real functions, $\lambda_1^\ast\left(x\right)=\lambda_2\left(x\right)$, and the operator on the left-hand side of Eq.~(\ref{eq1}) is positive definite.

Now we explain the difference in eigenstates between open and periodic boundary conditions. For example, we focus on the toy model described by Eq.~(\ref{eq1}) with
\begin{equation}
p\left(x\right)=p,~\lambda_1\left(x\right)=\lambda_2\left(x\right)=i\lambda\sin^2\frac{2\pi}{a}x,~v\left(x\right)=0,
\label{eq3}
\end{equation}
where $p$ and $\lambda$ are real constants. Importantly, the imaginary gauge potential represented by $\lambda_{1,2}\left(x\right)$ causes the non-Hermitian skin effect. We show the eigenvalues of the system with size $L$ under open and periodic boundary conditions in Fig.~\ref{fig1}(a). Surprisingly, the bands under the periodic boundary condition and those under the open boundary condition are different even in the limit of a large system size, in contrast to Hermitian systems, where they become asymptotically identical. The difference is caused by the non-Hermitian skin effect. Namely, under a periodic boundary condition, the eigenstates in the bulk extend over the system [Fig.~\ref{fig1}(b), top panel]. On the other hand, under an open boundary condition, the non-Hermitian skin effect occurs, and the bulk eigenstates are localized at either end of the system [Fig.~\ref{fig1}(b), bottom panel].

In Fig.~\ref{fig2}(a), we show the spatial distribution of the eigenstates with the Dirichlet boundary condition, $\psi\left(0\right)=\psi\left(L\right)=0$, and those with the Neumann boundary condition, $\psi^\prime\left(0\right)=\psi^\prime\left(L\right)=0$. Obviously, these eigenstates depend on the boundary conditions. Nevertheless, it is surprising that in the limit of a large system size, the asymptotic bands under the Dirichlet boundary condition are identical to those under the Neumann boundary condition [Fig.~\ref{fig2}(b)].

\begin{figure*}[ptb]
\includegraphics[clip,width=0.9\textwidth]{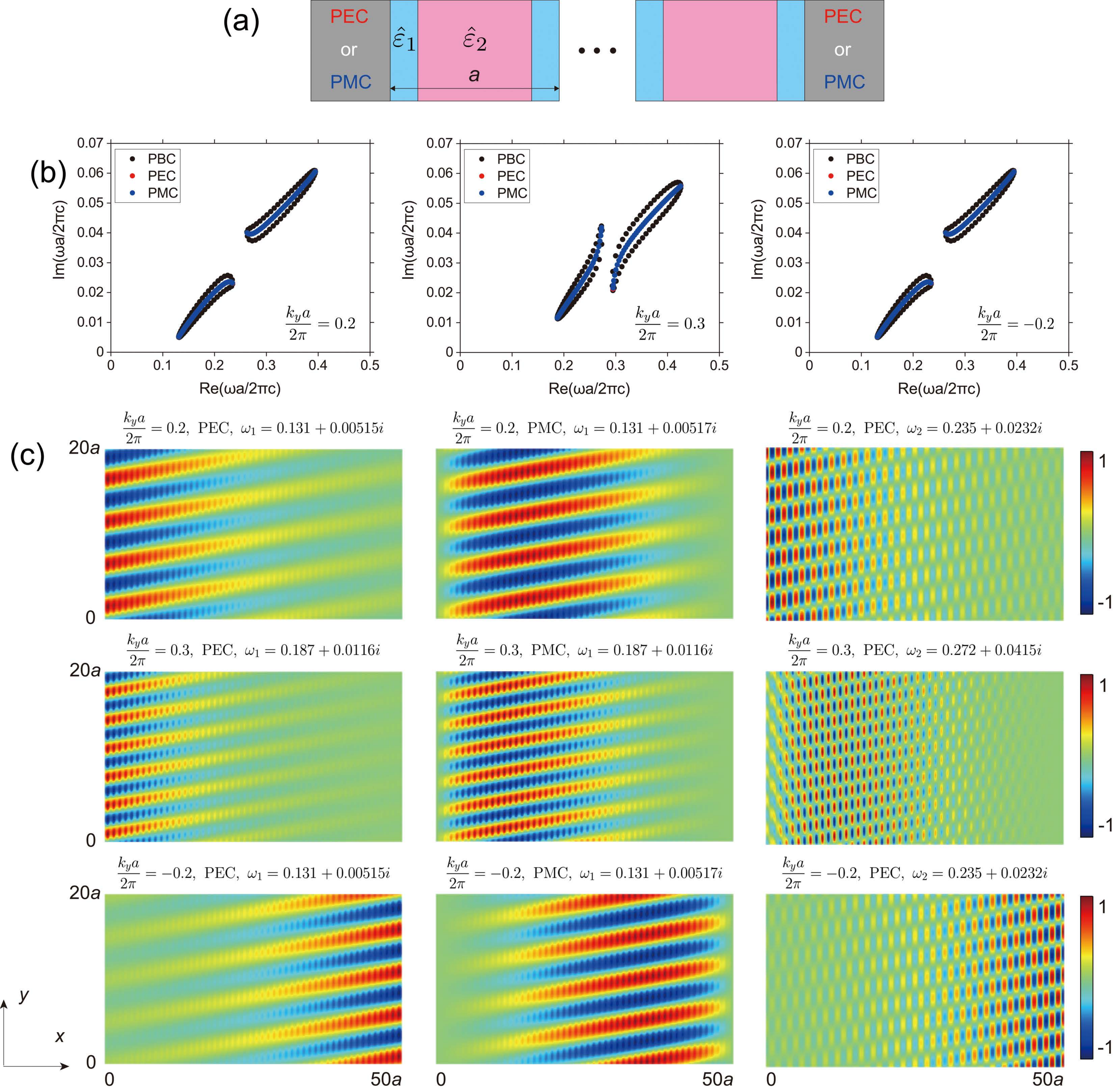}
\caption{\label{fig3}
(a) Photonic crystal with a system size of $50a$ under the perfect electric conductor (PEC) boundary condition and under the perfect magnetic conductor (PMC) boundary condition, where $a$ is the lattice constant. We set the system parameters as $\varepsilon_{1,xx}=\varepsilon_{1,yy}=1,\varepsilon_{1,xy}=\varepsilon_{yx}=0,\varepsilon_{2,xx}=\varepsilon_{2,yy}=9+3i,\varepsilon_{2,xy}=\varepsilon_{2,yx}=2,d_1=0.4a$, and $d_2=0.6a$.
(b) Eigenvalues with the PEC boundary condition (red), those with the PMC boundary condition (blue), and those with a periodic boundary condition (black). The red and blue dots overlap each other.
(c) Spatial distributions of the eigenstates under the PEC boundary condition and those under the PMC boundary condition for various values of $k_y$. In the left and middle columns, the eigenstates have the minimum value of the real part of the eigenvalues included in the first band, denoted as $\omega_1$. In the right column, the eigenstates have the maximum value of the real part of the eigenvalues included in the first band, denoted as $\omega_2$. We show the real part of the magnetic field.}
\end{figure*}

As another example, we focus on a photonic crystal, as shown in Fig.~\ref{fig3}(a). It is a multilayer in which two dielectric media are alternately stacked along the $x$ direction, and it is uniform in the other directions. To include an effective gauge potential for electromagnetic waves, we introduce anisotropy of the dielectric tensor~\cite{Liu2015,Liu2015v2,Chen2019}. Thus, we assume that the dielectric media with thickness $d_i~(i=1,2)$ have dielectric tensors
\begin{eqnarray}
\hat{\varepsilon}_i=\left( \begin{array}{ccc}
\varepsilon_{i,xx} & \varepsilon_{i,xy} & 0           \vspace{3pt}\\
\varepsilon_{i,yx} & \varepsilon_{i,yy} & 0           \vspace{3pt}\\
0                  & 0                  & \varepsilon
\end{array}\right)~\left(i=1,2\right)
\label{eq4}
\end{eqnarray}
and permeability $\mu_i=1~(i=1,2)$. We note that a real part of permittivity means anisotropy of the phase velocity, and an imaginary part of permittivity means anisotropy of the optical gain and loss. Throughout this paper, we investigate the reciprocal photonic crystal with $\hat{\varepsilon}_i^{\rm T}=\hat{\varepsilon}_i~(i=1,2)$. The lattice constant is $d_1+d_2\equiv a$. The electromagnetic waves in the multilayer with the dielectric tensor $\hat{\varepsilon}\left(x\right)$ are described by Maxwell's equations. Importantly, when the electromagnetic waves propagate in the $xy$ plane, transverse-electric (TE) modes are decoupled from transverse-magnetic (TM) modes because $\varepsilon_{xz},\varepsilon_{zx},\varepsilon_{yz},\varepsilon_{zy}=0$. In particular, for the TE modes, the governing equation can be written in the form of Eq.~(\ref{eq1}). In fact, by expressing the $z$ component of the magnetic field as $H_z\left(x,y\right)=H\left(x\right)e^{ik_yy}$, we can get the wave equation:
\begin{eqnarray}
&&\Biggl\{-\frac{d}{dx}\eta_{yy}\left(x\right)\frac{d}{dx} \nonumber\\
&&-\frac{i}{2}\left[-2k_y\eta_{xy}\left(x\right)\frac{d}{dx}+\frac{d}{dx}\left(-2k_y\eta_{yx}\left(x\right)\right)\right] \nonumber\\
&&+k_y^2\eta_{xx}\left(x\right)\Biggr\}H\left(x\right)=\left(\frac{\omega}{c}\right)^2H\left(x\right),
\label{eq5}
\end{eqnarray}
where $c$ is the speed of light in vacuum, and $\eta_{ij}\left(x\right)~(i,j=x,y)$ are the components of the inverse of $\hat{\varepsilon}\left(x\right)$. We explain the detailed derivation of Eq.~(\ref{eq5}) in Appendix~\ref{secA}. When $\hat{\varepsilon}^\dag\left(x\right)\neq\hat{\varepsilon}\left(x\right)$, the system becomes non-Hermitian. In particular, when $\eta_{xy}\left(x\right)$ and $\eta_{yx}\left(x\right)$ take complex values, the terms including the factors $-2k_y\eta_{xy}\left(x\right)$ and $-2k_y\eta_{yx}\left(x\right)$ in Eq.~(\ref{eq5}) express the imaginary gauge potential, and these terms give rise to the non-Hermitian skin effect.

In the system with size $L$ along the $x$ axis, by using COMSOL MULTIPHYSICS, we numerically calculate eigenvalues and eigenstates under the perfect electric conductor (PEC) boundary condition and those under the perfect magnetic conductor (PMC) boundary condition~\cite{Jackson1999}. We note that the PEC boundary condition and the PMC boundary condition are given by $E_y\left(0\right)=E_y\left(L\right)=0$ and by $H_z\left(0\right)=H_z\left(L\right)=0$, respectively. In Fig.~\ref{fig3}(b), as a manifestation of the non-Hermitian skin effect, the bands in the open boundary conditions and those in a periodic boundary condition are different even in the limit of a large system size, similar to the toy model. In fact, for some eigenvalues, the eigenstates are localized at the boundaries as shown in Fig.~\ref{fig3}(c). Surprisingly, the bands with the PEC boundary condition and those with the PMC boundary condition are identical [Fig.~\ref{fig3}(b)] although their eigenstates are quite different [Fig.~\ref{fig3}(c)]. It is worth noting that these states are not discrete edge states but continuum bulk eigenstates. In fact, at the boundaries, the density of these localized states is different from that of discrete edge states.

We comment on the numerical simulation by using COMSOL MULTIPHYSICS. We discuss the time evolution of the waves for the factor $e^{-i\omega t}$ throughout this paper. On the other hand, in COMSOL MULTIPHYSICS, the convention of the time evolution of the electromagnetic waves for the factor $e^{i\omega t}$ is adopted. To incorporate this difference in the convention, in the simulation by COMSOL MULTIPHYSICS, we take the input parameters to be complex conjugate of the original ones; that is, in the simulation, we input $\varepsilon_{2,xx}=\varepsilon_{2,yy}=9-3i$ instead of $9+3i$.

Thus the non-Hermitian skin effect causes various remarkable phenomena. In our work, we show that these phenomena can be understood in terms of the non-Bloch band theory.

%
%

\section{\label{sec3}Non-Bloch band theory}
Eigenstates under a periodic boundary condition are described in terms of a real wave number, stemming from translational symmetry. In contrast, under an open boundary condition, translational symmetry is violated, and such cases are outside of the conventional Bloch theory. Nevertheless, supposing that such cases can be described in terms of a Bloch wave number, the non-Hermitian skin effect implies that it takes complex values. Indeed, the non-Bloch band theory proposed in our previous works~\cite{Yokomizo2019,Yokomizo2020,Yokomizo2021} shows that non-Hermitian periodic systems are described in terms of the complex Bloch wave number. Now we explain the concept of the non-Bloch band theory. To this end, we focus on a non-Hermitian tight-binding system with a finite system size. Here this system has discrete eigenvalues. Then, as the system size increases, the eigenvalues become dense. Finally, in the limit of a large system size, the discrete eigenvalues form continuous sets, being eigenspectra. From our previous works, the eigenspectra are calculated from sets of the complex values of $\beta=e^{ika}$, where $k$ is the complex Bloch wave number. Each set of the value of $\beta$ forms a loop on the complex plane, and it is called the generalized Brillouin zone. We note that these loops become a unit circle under a periodic boundary condition. The non-Bloch band theory provides a method to calculate the generalized Brillouin zone reproducing eigenspectra. We briefly explain a way to get the generalized Brillouin zone in Appendix~\ref{secB}.

\begin{figure*}[ptb]
\includegraphics[clip,width=0.9\textwidth]{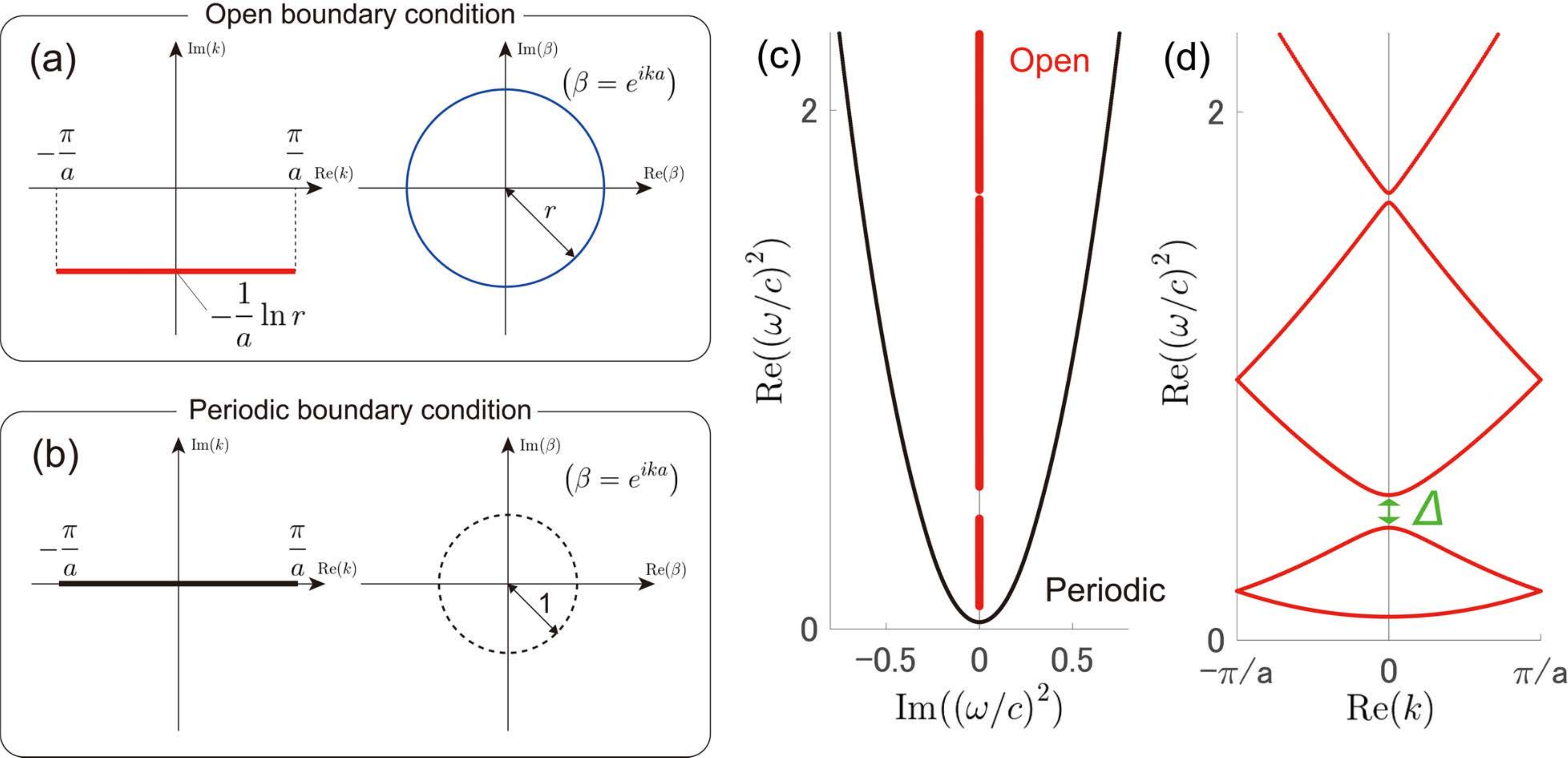}
\caption{\label{fig4}
(a) Complex Bloch wave number and generalized Brillouin zone. The radius $r$ is given in Eq.~(\ref{eq6}).
(b) Real Bloch wave number and Brillouin zone.
(c) Eigenspectra from the generalized Brillouin zone (red) and those from the conventional Brillouin zone (black) in the toy model (\ref{eq3}). The system parameters are set to be $p=10^{-2}$ and $\lambda=10^{-1}$.
(d) Reduced zone representation of the eigenspectra from the generalized Brillouin zone. The gap $\Delta$ is given in Eq.~(\ref{eq10}).}
\end{figure*}

In the present work, we extend the idea of the non-Bloch band theory to a non-Hermitian continuous system described by Eq.~(\ref{eq1}). Then we derive a formula of the generalized Brillouin zone by using the non-Bloch band theory. When we discretize the system by dividing the unit cell into $N$ equal parts, the operator on the left-hand side of Eq.~(\ref{eq1}) is approximated by an $N\times N$ matrix (see Appendix~\ref{secC}). Since the non-Bloch band theory can be applied to this discrete system, we can get the formula for the generalized Brillouin zone. Finally, in the limit of $N\rightarrow\infty$, we show that in the continuous system, the generalized Brillouin zone becomes a circle with the radius
\begin{equation}
r=\exp\left(\frac{1}{2}\int_0^adx\Im\frac{\lambda_1\left(x\right)+\lambda_2\left(x\right)}{2p\left(x\right)}\right).
\label{eq6}
\end{equation}
We explain a way to derive Eq.~(\ref{eq6}) in Appendix~\ref{secC}. In this case, the imaginary part of the Bloch wave number $\Im\left(k\right)$ is obtained by $-\frac{1}{a}\ln r$. We note that Eq.~(\ref{eq6}) is independent of the complex potential $v\left(x\right)$, although it leads to non-Hermiticity. In Fig.~\ref{fig4}(a), we show the complex Bloch wave number and the generalized Brillouin zone which corresponds to an open boundary condition. In Fig.~\ref{fig4}(b), we also show the real Bloch wave number and the conventional Brillouin zone.

From Eq.~(\ref{eq6}), we can calculate the eigenspectra of Eq.~(\ref{eq1}). Because of the spatial periodicity (\ref{eq2}), the solution of Eq.~(\ref{eq1}) can be written in the form of the plane wave expansion:
\begin{equation}
\psi\left(x\right)=\sum_n\Psi\left(k+\frac{2n\pi}{a}\right)e^{i\left(k+\frac{2n\pi}{a}\right)x},
\label{eq7}
\end{equation}
where $k$ is the complex Bloch wave number on the generalized Brillouin zone. Then we can derive the secular equation from Eq.~(\ref{eq1}) as
\begin{eqnarray}
&&\left(\frac{\omega}{c}\right)^2\Psi\left(k+\frac{2n\pi}{a}\right)-\sum_{n^\prime}\Biggl\{\left(k+\frac{2n\pi}{a}\right)P_{n-n^\prime}\left(k+\frac{2n^\prime\pi}{a}\right) \nonumber\\
&&+\frac{1}{2}\left[\Lambda_{1,n-n^\prime}\left(k+\frac{2n^\prime\pi}{a}\right)+\left(k+\frac{2n\pi}{a}\right)\Lambda_{2,n-n^\prime}\right] \nonumber\\
&&+V_{n-n^\prime}\Biggr\}\Psi\left(k+\frac{2n^\prime\pi}{a}\right)=0,
\label{eq8}
\end{eqnarray}
where $P_n,\Lambda_{i,n}~(i=1,2)$, and $V_n$ are the Fourier coefficients of the functions $p\left(x\right),\lambda_i\left(x\right)~(i=1,2)$, and $v\left(x\right)$, respectively. Therefore, by combining the generalized Brillouin zone with Eq.~(\ref{eq8}), we can get the eigenspectra of the system.

%
%

\section{\label{sec4}Examples}

%
%

\subsection{\label{sec4-1}Toy model}
In the toy model (\ref{eq3}), we calculate the generalized Brillouin zone and the eigenspectra in order to check that it really reproduces the eigenvalues under an open boundary condition. From Eq.~(\ref{eq6}), the generalized Brillouin zone is a circle with the radius given by
\begin{equation}
r=\exp\left(-\frac{a\lambda}{4p}\right).
\label{eq9}
\end{equation}
As a result, from Eqs.~(\ref{eq8}) and (\ref{eq9}), the eigenspectra can be calculated as shown in Fig.~\ref{fig4}(c). In Fig.~\ref{fig4}(c), we also show the eigenspectra of the system obtained from the conventional Brillouin zone. Compared with Fig.~\ref{fig1}(a), we can confirm that our analytic calculation matches the results in the finite system. In fact, in the limit of a large system size, both the eigenvalues under the Dirichlet and Neumann boundary conditions asymptotically become identical with the eigenspectra from the generalized Brillouin zone, as shown in Fig.~\ref{fig2}(b). Thus, the eigenvalues are independent of the type of open boundary conditions, although the corresponding eigenstates are different [Fig.~\ref{fig2}(a)]. This counterintuitive conclusion can be obtained from the fact that the generalized Brillouin zone is independent of boundary conditions of an open system.

We note that in a tight-binding model, a complex eigenspectrum in a periodic chain forms a closed loop on the complex plane, and it surrounds eigenspectra in an open chain in general~\cite{Okuma2020,Zhang2020}. In contrast, we find that the complex eigenspectrum obtained from the conventional Brillouin zone does not form a loop, and it does not surround the eigenspectra obtained from the generalized Brillouin zone, which never occurs in non-Hermitian discrete systems. Such a distribution of the eigenspectrum is unique to continuous systems~\cite{Longhi2021}.

Finally, we show the reduced zone representation of the eigenspectra from the generalized Brillouin zone in Fig.~\ref{fig4}(d). Then, as an example, we analytically estimate the gap $\Delta$ between the second and third bands at $\Re\left(k\right)=0$, as
\begin{equation}
\Delta=\frac{\lambda^2}{8p}.
\label{eq10}
\end{equation}
We derive Eq.~(\ref{eq10}) in detail in Appendix~\ref{secD}. With our parameters, the value of Eq.~(\ref{eq10}) matches the numerical result.

%
%

\subsection{\label{4-2}Photonic crystal with anisotropy and loss}
\begin{figure*}[ptb]
\includegraphics[clip,width=0.9\textwidth]{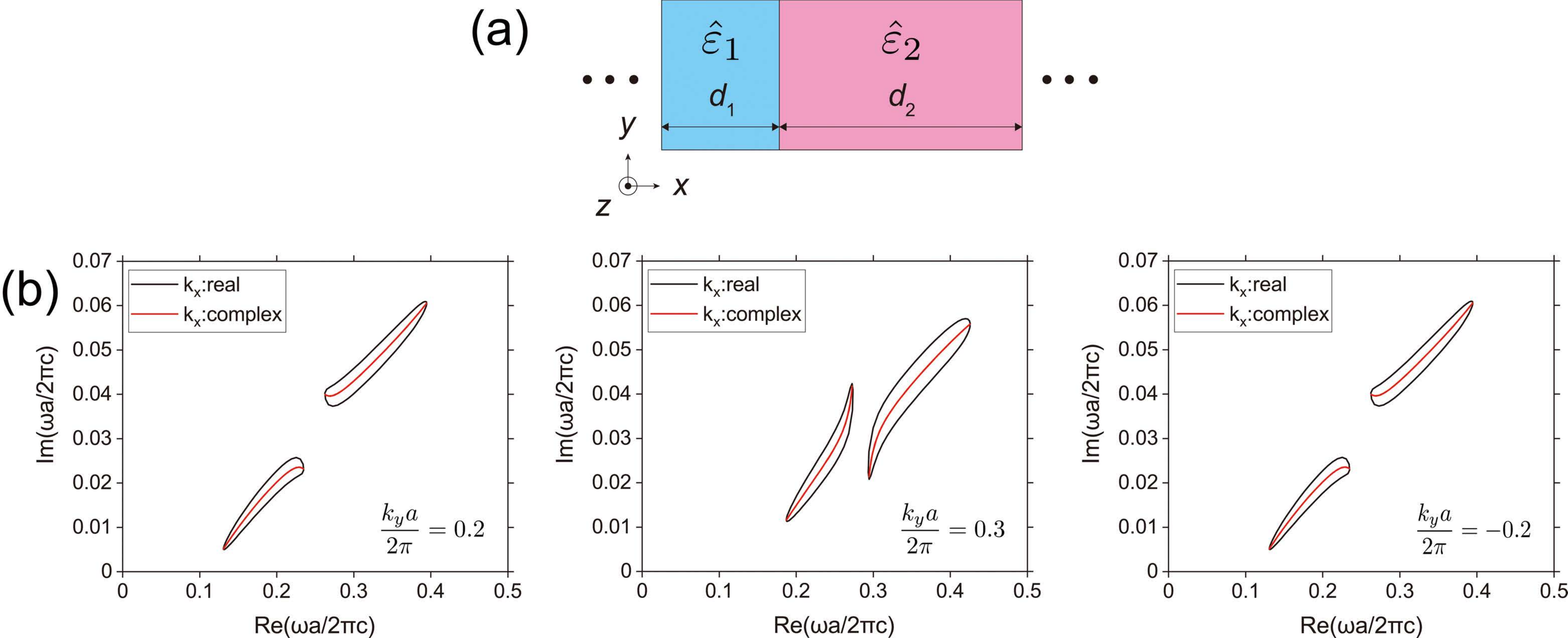}
\caption{\label{fig5}
(a) Photonic crystal. The photonic crystal is an alternate stacking of two media along the $x$ direction, and it is extended along the $y$ and $z$ directions. The component of the dielectric tensors are set to be $\varepsilon_{1,xx}=\varepsilon_{1,yy}=1,\varepsilon_{1,xy}=\varepsilon_{yx}=0,\varepsilon_{2,xx}=\varepsilon_{2,yy}=9+3i$, and $\varepsilon_{2,xy}=\varepsilon_{2,yx}=2$, and the thickness of the layer is set to be $d_1=0.4a$ and $d_2=0.6a$, where $a$ is a lattice constant.
(b) Eigenspectra from the generalized Brillouin zone (red) and those from the conventional Brillouin zone (black) for various values of $k_y$.}
\end{figure*}
In terms of the non-Bloch band theory, we can understand the non-Hermitian physics of the photonic crystal as shown in Fig.~\ref{fig5}(a). As explained above, the Bloch wave number $k_x$ becomes complex in the system. In fact, from Eq.~(\ref{eq6}), the generalized Brillouin zone becomes a circle with the radius
\begin{equation}
r=\exp\left(\frac{k_y}{2}\Im\sum_{i=1}^2\frac{\varepsilon_{i,xy}+\varepsilon_{i,yx}}{\varepsilon_{i,xx}}d_i\right).
\label{eq11}
\end{equation}
Importantly, the non-Hermitian skin effect occurs even if the dielectric tensor is symmetric. This means that the violation of the Lorentz reciprocity is not necessarily required~\cite{Jalas2013}. We note that Eq.~(\ref{eq11}) can also be derived by using a transfer matrix. In Appendix~\ref{secE}, we show that the transfer matrix can give the generalized Brillouin zone.

In Fig.~\ref{fig5}(b), we show the eigenspectra from the generalized Brillouin zone with the complex $k_x$ and those from the conventional Brillouin zone with the real $k_x$ for various values of $k_y$. We can confirm that these spectra are different from each other. Importantly, the eigenspectra calculated from Eq.~(\ref{eq11}) are expected to reproduce eigenvalues with an open boundary condition in the limit of a large system size, regardless of the details of open boundary conditions. In fact, in comparison with Figs.~\ref{fig3}(b) and \ref{fig5}(b), the eigenvalues under the PEC boundary condition and those under the PMC boundary condition match the eigenspectra. Thus, the asymptotic behavior of the system in the limit of a large system size is independent of the type of open boundary conditions. Remarkably, with the constant $k_y$, the localization length of all the eigenstates is common. This reflects that the generalized Brillouin zone is a circle. The localization length of the eigenstates is determined by the imaginary part of the complex Bloch wave number, which is given by $\Im\left(k_x\right)=0.04k_y$ in our computation. Thus, for $\frac{k_ya}{2\pi}=0.2,0.3,-0.2$, we get $\Im\left(k_x\right)a=0.016\pi,0.024\pi,-0.016\pi$, respectively. This means that for $k_y>0$ ($k_y<0$), the eigenstates are localized at the left (right) end of the system and the localization lengths are $\frac{1}{\left|\Im\left(k_x\right)\right|}\simeq20a,13a,20a$ for the above three cases. The conclusion obtained here is consistent with the non-Bloch band theory, as mentioned above.

%
%

\section{\label{sec5}Bulk-edge correspondence}
\begin{figure*}[ptb]
\includegraphics[clip,width=0.9\textwidth]{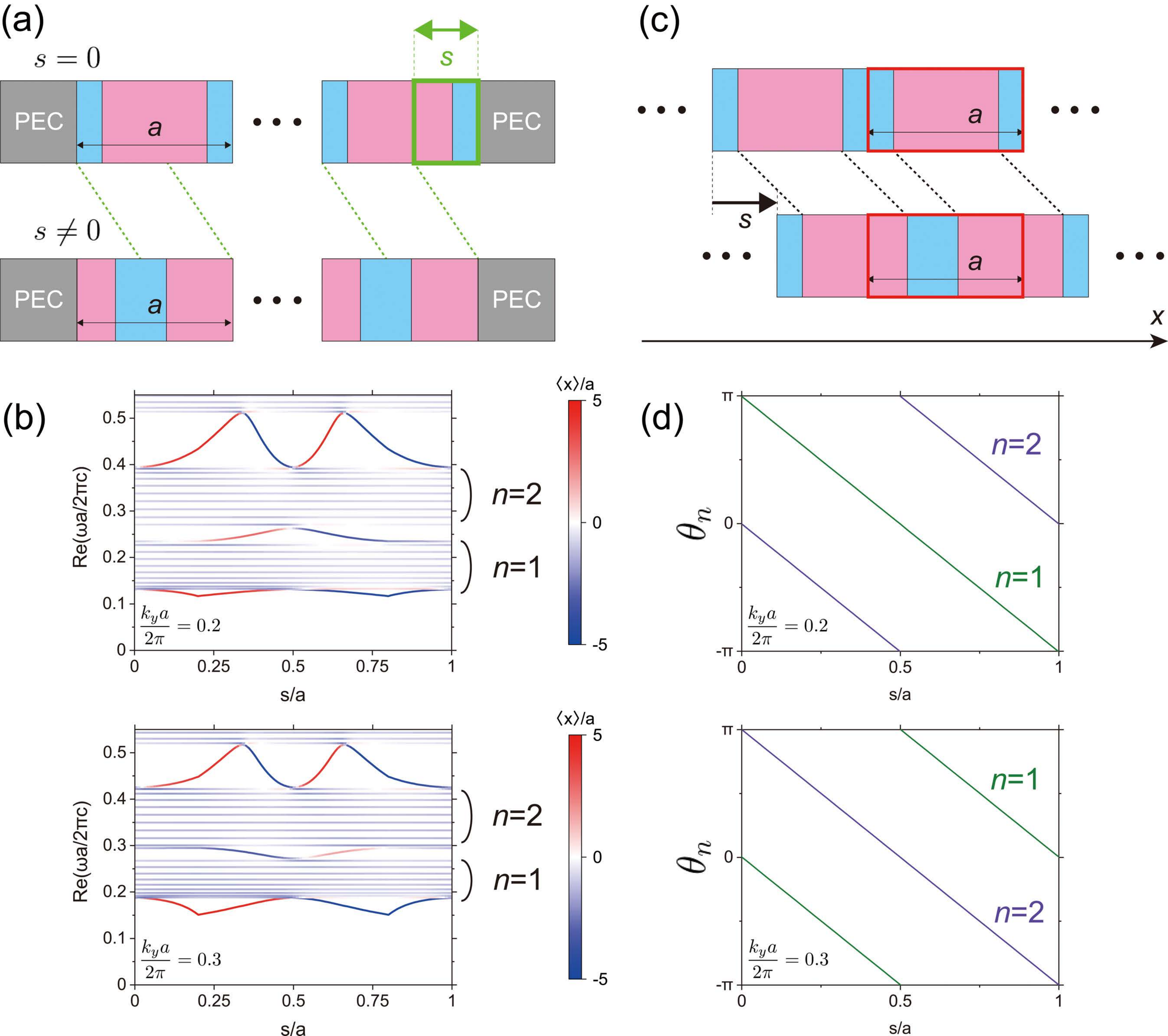}
\caption{\label{fig6}
(a) Change in the photonic crystal via the change in the parameter $s$. The photonic crystal has a system size of $10a$ with the perfect electric conductor (PEC) boundary condition. The system parameters are the same as in Figs.~\ref{fig3} and \ref{fig5}.
(b) Eigenvalues as a function of $s$. We also show the edge states localized at the right boundary (red) and those localized at the left boundary (blue). $n$ expresses the band index.
(c) Configuration of the unit cell (red) for the spatial translation parameter $s$.
(d) $s$ dependency of the Zak phase $\theta_n$.}
\end{figure*}
In non-Hermitian systems, the bulk-edge correspondence between a topological invariant defined in terms of the conventional Brillouin zone and the appearance of topological edge states was thought to be violated~\cite{Lee2016}. This is because the eigenvalues with an open boundary condition and those with a periodic boundary condition are different. Hence, eigenspectra calculated from the conventional Brillouin zone cannot predict a bulk-gap-closing point and topological phase transition under an open boundary condition. Thus, it is necessary to discuss the bulk-edge correspondence in terms of the non-Bloch band theory. Now we demonstrate the bulk-edge correspondence in the photonic crystal by using the topological invariant defined from the generalized Brillouin zone.

First of all, we focus on the photonic crystal with a finite system size under the PEC boundary condition. Then, as shown in Fig.~\ref{fig6}(a), we change the termination of the system so that we cut off the layer at the right end by the thickness $s$ and add to the layer at the left end by the same thickness. In this case, the system size is unchanged. With the system size $L=10a$, we show the spectral flow through the change in $s$ and the center of position, defined as
\begin{equation}
\left<x\right>=\frac{\displaystyle\int_{-L/2}^{L/2}dxx\left|H\left(x\right)\right|^2}{\displaystyle\int_{-L/2}^{L/2}dx\left|H\left(x\right)\right|^2},
\label{eq12}
\end{equation}
in Fig.~\ref{fig6}(b). We find that $2n$ edge states appear in the band gap between the $n$th bulk band and the $\left(n+1\right)$th bulk band. Furthermore, at the $n$th gap, $n$ states are localized at the right boundary, and the other $n$ states are localized at the left boundary.

As we change the termination in the finite system as shown in Fig.~\ref{fig6}(a), we change the unit cell in the bulk, so that the boundaries of the unit cell match those of the whole system, as shown in Fig.~\ref{fig6}(c). Hence, the shift of the unit cell can be represented by the parameter $s$. Then, according to \cite{Nakata2020}, the appearance of the edge states corresponds to the change in the Zak phase through the change in the unit cell. In our work, the Zak phase $\theta_n$ can be defined in terms of the generalized Brillouin zone (see Appendix~\ref{secF}). We note that the photonic crystal belongs to class A in a $k_x$-$s$ plane, where $k_x$ is the complex wave number on the generalized Brillouin zone. In fact, the integral of the Zak phase for the variable $s$ is equivalent to the Chern number defined in a $k_x$-$s$ plane. In Fig.~\ref{fig6}(d), we numerically calculate the $s$ dependency of the Zak phase. Thus, the values of the Zak phase change by $-2\pi$ under the change in $s$ from 0 to 1. We note that at $\frac{s}{a}=0,0.5,1$, the values of the Zak phase are close to 0 and $\pm\pi$, but they are not quantized. This is because the off-diagonal components of the dielectric tensor break the mirror symmetry in the $x$ direction. Our numerical results indicate that one mode is carried from the left end to the right end per band. Thus, within the $n$th gap, $n$ states go out from the $n$th band at the right end, and $n$ states enter the $n$th band at the left end. Therefore, the existence of the edge states coincides with the nontrivial topological invariant, and we conclude that the edge states are topological.

%
%

\section{\label{sec6}Summary and Discussion}
In this work, we studied the non-Hermitian waves described in the wave equation (\ref{eq1}) with a continuous periodic model in terms of the non-Bloch band theory. As examples, we investigated the toy model given in Eq.~(\ref{eq3}) and the photonic crystal shown in Fig.~\ref{fig5}(a) and showed that the bulk eigenstates in both systems exhibit the non-Hermitian skin effect. We found that the asymptotic behavior of the system in the limit of a large system size is independent of the type of open boundary conditions, such as the Dirichlet and Neumann boundary conditions in the toy model and the PEC and PMC boundary conditions in the photonic crystal. In fact, the asymptotic eigenvalues become identical with the eigenspectra calculated from the generalized Brillouin zone. Remarkably, this means that the localization length for all the skin modes is common because the generalized Brillouin zone becomes a unit circle with the radius given in Eq.~(\ref{eq6}). Furthermore, the generalized Brillouin zone gives various physical properties in non-Hermitian systems. For example, we can establish the bulk-edge correspondence between the Zak phase defined from the generalized Brillouin zone and the existence of the topological edge states.

The non-Hermitian skin effect in a continuous system was studied in terms of non-Hermitian topology in Refs.~\cite{Longhi2021,Zhong2021}. These previous works showed that in a continuous system, the non-Hermitian skin effect corresponds to the energy winding number. In particular, Ref.~\cite{Longhi2021} investigated the energy winding number in terms of the imaginary gauge transformation. The imaginary gauge transformation reveals detailed properties of the non-Hermitian skin effect, such as the localization length. Nevertheless, the imaginary gauge transformation discussed in Ref.~\cite{Longhi2021} is limited to some specific models. On the other hand, our work revealed that the non-Bloch band theory is applicable to general Strum-Liouville-type non-Hermitian continuous systems, where the imaginary gauge transformation may not be applicable. In fact, we found that all the eigenstates have a common localization length in the system described by Eq.~(\ref{eq1}) due to the circular shape of the generalized Brillouin zone. Thus, the remarkable features of the non-Hermitian skin effect are expected to be present in various continuous systems, such as an elastic medium, an acoustic medium, and a photonic crystal.

In this paper, we established the property that the localization lengths of the skin modes are independent of eigenvalues in a non-Hermitian system described by a second-order differential equation. Since various physical systems are described in terms of the Sturm-Liouville equation, this conclusion is applicable to a wide range of physical systems. Whether this property holds in arbitrary non-Hermitian continuous systems is left as future work.

Anisotropy and loss of a dielectric medium can be readily achieved by natural hyperbolic materials~\cite{Narimanov2015,Korzeb2015} and hyperbolic metamaterials~\cite{Poddubny2013,Kildishev2013,Jahani2016}. Hence, we expect that the photonic crystal discussed here can be experimentally realized. In addition, dynamical Floquet modulation could be another candidate for realizing the imaginary gauge potential~\cite{He2019,Fang2019,Lu2021}. One can find a wide variety of possible realizations because one-dimensional propagation of electromagnetic waves can be described by the wave equation (\ref{eq1}) in general. Interestingly, we can confine electromagnetic waves in a photonic crystal by using anisotropy and non-Hermiticity of a dielectric medium. Finally, theoretical and experimental studies on the propagation of electromagnetic waves in two-dimensional and three-dimensional non-Hermitian systems are left as future works.

%
%

\begin{acknowledgements}
K.Y. and S.M. are grateful to T. Matsuo for valuable discussion. K.Y. is also grateful to T. Sasamoto, R. Takahashi, and Y. Takahashi. T.Y. thanks Y. Moritake, K. Takata, and M. Notomi for helpful discussions. This work was supported by JSPS KAKENHI (Grant No.~JP18H03678) and by the MEXT Elements Strategy Initiative to Form Core Research Center (TIES; Grant No.~JPMXP0112101001). K.Y. was also supported by JSPS KAKENHI Grant No.~JP21J01409.
\end{acknowledgements}

%
%

\appendix

%
%

\section{\label{secA}Wave equation in the photonic crystal}
We show that Maxwell's equations can be written in the form of Eq.~(\ref{eq5}). In the following, we study a multilayer system composed of two dielectric media, as shown in Fig.~\ref{fig5}(a). The dielectric tensor of the multilayer is given by
\begin{eqnarray}
\begin{array}{l}
\hat{\varepsilon}\left(x\right)=\left\{ \begin{array}{ll}
\hat{\varepsilon}_1 & \left(0\leq x\leq d_1\right), \vspace{3pt}\\
\hat{\varepsilon}_2 & \left(d_1\leq x\leq a\right),
\end{array}\right. \vspace{3pt}\\
\hat{\varepsilon}\left(x+a\right)=\hat{\varepsilon}\left(x\right),
\end{array}
\label{eqapp11}
\end{eqnarray}
where
\begin{eqnarray}
\hat{\varepsilon}_i=\left( \begin{array}{ccc}
\varepsilon_{i,xx} & \varepsilon_{i,xy} & 0           \vspace{3pt}\\
\varepsilon_{i,yx} & \varepsilon_{i,yy} & 0           \vspace{3pt}\\
0                  & 0                  & \varepsilon
\end{array}\right)~\left(i=1,2\right)
\label{eqapp12}
\end{eqnarray}
and $a=d_1+d_2$ is a lattice constant. We assume that the permeability is $1$ throughout this multilayer. Then Maxwell's equations are written as
\begin{eqnarray}
\left\{ \begin{array}{l}
{\bm\nabla}\cdot{\bm H}\left({\bm r},t\right)=0, \vspace{5pt}\\
{\bm\nabla}\cdot\left(\hat{\varepsilon}\left(x\right){\bm E}\left({\bm r},t\right)\right)=0, \vspace{5pt}\\
\displaystyle{\bm\nabla}\times{\bm E}\left({\bm r},t\right)+\mu_0\frac{\partial}{\partial t}{\bm H}\left({\bm r},t\right)={\bm0}, \vspace{5pt}\\
\displaystyle{\bm\nabla}\times{\bm H}\left({\bm r},t\right)-\varepsilon_0\hat{\varepsilon}\left(x\right)\frac{\partial}{\partial t}{\bm E}\left({\bm r},t\right)={\bm 0},
\end{array}\right.
\label{eqapp13}
\end{eqnarray}
where $\varepsilon_0$ and $\mu_0$ are the vacuum permittivity and the vacuum permeability, respectively. In our work, we study the electromagnetic wave with frequency $\omega$ given by
\begin{equation}
{\bm H}\left({\bm r},t\right)=\tilde{\bm H}\left({\bm r}\right)e^{-i\omega t},~{\bm E}\left({\bm r},t\right)=\tilde{\bm E}\left({\bm r}\right)e^{-i\omega t}.
\label{eqapp14}
\end{equation}
In this case, Eq.~(\ref{eqapp13}) can be rewritten in the form of an eigenvalue equation~\cite{Joannopoulos2008}:
\begin{equation}
{\bm\nabla}\times\left(\frac{1}{\hat{\varepsilon}\left(x\right)}{\bm\nabla}\times\tilde{\bm H}\left({\bm r}\right)\right)=\left(\frac{\omega}{c}\right)^2\tilde{\bm H}\left({\bm r}\right),
\label{eqapp15}
\end{equation}
where $c$ is the speed of light in vacuum. For convenience, let us denote $\frac{1}{\hat{\varepsilon}\left(x\right)}$ as
\begin{eqnarray}
\frac{1}{\hat{\varepsilon}\left(x\right)}\equiv\eta\left(x\right)=\left( \begin{array}{ccc}
\eta_{xx}\left(x\right) & \eta_{xy}\left(x\right) & 0    \vspace{3pt}\\
\eta_{yx}\left(x\right) & \eta_{yy}\left(x\right) & 0    \vspace{3pt}\\
0                       & 0                       & \eta
\end{array}\right).
\label{eqapp16}
\end{eqnarray}
Now we focus on the transverse-electric (TE) modes propagating in the $xy$ plane in the multilayer, which means that the electromagnetic waves are independent of the $z$ coordinate. In this case, the TE modes have a magnetic field normal to the $xy$ plane, $\tilde{\bm H}\left(x,y\right)=\tilde{H}_z\left(x,y\right)\hat{\bm z}$, and the electric field in the $xy$ plane, $\tilde{\bm E}\left(x,y\right)\cdot\hat{\bm z}=0$. Furthermore, we can rewrite the magnetic field as $\tilde{H}_z\left(x,y\right)=H_0\left(x\right)e^{ik_yy}$. Finally, for $H_0\left(x\right)$, Eq.~(\ref{eqapp15}) can be explicitly written as
\begin{eqnarray}
&&\Biggl\{-\frac{d}{dx}\eta_{yy}\left(x\right)\frac{d}{dx} \nonumber\\
&&-\frac{i}{2}\left[-2k_y\eta_{xy}\left(x\right)\frac{d}{dx}+\frac{d}{dx}\left(-2k_y\eta_{yx}\left(x\right)\right)\right] \nonumber\\
&&+k_y^2\eta_{xx}\left(x\right)\Biggr\}H_0\left(x\right)=\left(\frac{\omega}{c}\right)^2H_0\left(x\right).
\label{eqapp17}
\end{eqnarray}
Thus, we can derive Eq.~(\ref{eq5}). This equation corresponds to Eq.~(\ref{eq1}) with $p\left(x\right)=\eta_{yy}\left(x\right),\lambda_1\left(x\right)=-2k_y\eta_{xy}\left(x\right),\lambda_2\left(x\right)=-2k_y\eta_{yx}\left(x\right)$, and $v\left(x\right)=k_y^2\eta_{xx}\left(x\right)$. We note that the system becomes Hermitian when $\eta_{xx}\left(x\right),\eta_{yy}\left(x\right)\in{\mathbb R}$ and $\eta_{xy}^\ast\left(x\right)=\eta_{yx}\left(x\right)$. This is the case for $\hat{\varepsilon}^\dag\left(x\right)=\hat{\varepsilon}\left(x\right)$.

%
%

\section{\label{secB}Non-Bloch band theory}
We briefly review the non-Bloch band theory~\cite{Yokomizo2019}. To this end, we focus on a one-dimensional tight-binding system with an open boundary condition. The Hamiltonian of this system is written as
\begin{equation}
H=\sum_n\sum_{i=-N}^N\sum_{\mu,\nu=1}^qt_{i,\mu\nu}c_{n+i,\mu}^\dag c_{n,\nu},
\label{eqapp21}
\end{equation}
where $c_{n,\mu}^\dag$ is a creation operator of a particle on the $\mu$th sublattice in the $n$th unit cell. $N$ represents the hopping range of the particle, and $q$ is the number of internal degrees of freedom in the unit cell. When $t_{-i,\nu\mu}^\ast\neq t_{i,\mu\nu}$, the system becomes non-Hermitian. Now, for the eigenvector
\begin{equation}
|\psi\rangle=\left(\dots,\psi_{1,1},\dots\psi_{1,q},\dots,\psi_{L,1},\dots,\psi_{L,q},\dots\right)^{\rm T},
\label{eqapp22}
\end{equation}
the real-space eigen-equation is given by
\begin{equation}
H|\psi\rangle=E|\psi\rangle.
\label{eqapp23}
\end{equation}
Then we can get the solution of Eq.~(\ref{eqapp23}) as
\begin{equation}
\psi_{n,\mu}=\sum_j\left(\beta_j\right)^n\phi_\mu^{\left(j\right)}~\left(\mu=1,\dots,q\right)
\label{eqapp24}
\end{equation}
because of the spatial periodicity. Here $\beta=\beta_j$ is the solution of the characteristic equation
\begin{equation}
\det\left[{\cal H}\left(\beta\right)-E\right]=0,
\label{eqapp25}
\end{equation}
where ${\cal H}\left(\beta\right)$ is the non-Bloch matrix defined as
\begin{equation}
\left[{\cal H}\left(\beta\right)\right]_{\mu\nu}=\sum_{i=-N}^Nt_{i,\mu\nu}\beta^i~\left(\mu,\nu=1,\dots,q\right).
\label{eqapp26}
\end{equation}
In general, Eq.~(\ref{eqapp25}) is an algebraic equation with an even degree $2M=2qN$ for $\beta$. From the above, we can calculate the eigenvalues in a finite chain by combining Eq.~(\ref{eqapp25}) and an open boundary condition. In the limit of a large system size, the discrete eigenvalues form dense sets, and the eigenspectra can be obtained from the condition of the $2M$ solutions in Eq.~(\ref{eqapp25}), given by
\begin{equation}
\left|\beta_M\right|=\left|\beta_{M+1}\right|,
\label{eqapp27}
\end{equation}
with
\begin{equation}
\left|\beta_1\right|\leq\dots\leq\left|\beta_{2M}\right|.
\label{eqapp28}
\end{equation}
We note that the trajectories of $\beta_M$ and $\beta_{M+1}$ satisfying Eq.~(\ref{eqapp27}) are the generalized Brillouin zone. Therefore, Eq.~(\ref{eqapp27}) expresses the condition for the generalized Brillouin zone.

%
%

\section{\label{secC}Generalized Brillouin zone in continuous systems}
We explain a way to get the generalized Brillouin zone in a non-Hermitian continuous system described by the wave equation
\begin{eqnarray}
\left[-\frac{d}{dx}p\left(x\right)\frac{d}{dx}-\frac{i}{2}\left(\lambda_1\left(x\right)\frac{d}{dx}+\frac{d}{dx}\lambda_2\left(x\right)\right)+v\left(x\right)\right]\psi\left(x\right) \nonumber\\
=\left(\frac{\omega}{c}\right)^2\psi\left(x\right). \nonumber\\
\label{eqapp31}
\end{eqnarray}
We focus on a unit cell with the lattice constant $a$ in this system. First of all, we divide the unit cell into $N$ equal parts. The size of each part $\delta$ is given by $\frac{a}{N}$. In this case, the unit cell can be regarded as a tight-binding system with $N$ sites. Then the operator on the left-hand side of Eq.~(\ref{eqapp31}) can be approximated to the form of a matrix. In our work, we call this matrix the non-Bloch matrix. Importantly, it is necessary to discretize the operator by the central difference method. This is because when the system becomes Hermitian, the operator including $\lambda_1\left(x\right)$ and $\lambda_2\left(x\right)$ should have a skew symmetry. As a result, the non-Bloch matrix can be explicitly written as
\begin{eqnarray}
{\cal H}\left(\beta\right)=\left( \begin{array}{ccccc}
A_N      & B_1    &        &               & C_N\beta^{-1} \\
C_1      & A_1    & \ddots &               &               \\
         & \ddots & \ddots & \ddots        &               \\
         &        & \ddots & A_{N-2}       & B_{N-1}       \\
B_N\beta &        &        & C_{N-1}       & A_{N-1}
\end{array}\right),
\label{eqapp32}
\end{eqnarray}
where
\begin{eqnarray}
\left\{ \begin{array}{l}
\displaystyle A_j=\frac{p\left(x_j\right)+p\left(x_{j+1}\right)}{\delta^2}+v\left(x_j\right),                                   \vspace{5pt}\\
\displaystyle B_j=-\frac{p\left(x_j\right)}{\delta^2}-i\frac{\lambda_1\left(x_j\right)+\lambda_2\left(x_{j+1}\right)}{4\delta}, \vspace{5pt}\\
\displaystyle C_j=-\frac{p\left(x_j\right)}{\delta^2}+i\frac{\lambda_1\left(x_{j+1}\right)+\lambda_2\left(x_j\right)}{4\delta},
\end{array}\right.
\label{eqapp33}
\end{eqnarray}
for $j=1,\dots,N$, and $\beta=e^{ika}$. We note that $x_{N+1}=x_1$. Hence, we can get the characteristic equation of the non-Bloch matrix, written as
\begin{eqnarray}
&&\left(\frac{1}{\delta^2}\right)^N\prod_{j=1}^N\left[p\left(x_j\right)+\frac{i\delta}{4}\left(\lambda_1\left(x_j\right)+\lambda_2\left(x_{j+1}\right)\right)\right]\beta \nonumber\\
&&+\left(\frac{1}{\delta^2}\right)^N\prod_{j=1}^N\left[p\left(x_j\right)-\frac{i\delta}{4}\left(\lambda_1\left(x_{j+1}\right)+\lambda_2\left(x_j\right)\right)\right]\beta^{-1} \nonumber\\
&&+\left(\beta~{\rm independent~term}\right)=0.
\label{eqapp34}
\end{eqnarray}
Since Eq.~(\ref{eqapp34}) is a quadratic equation for $\beta$, from Eq.~(\ref{eqapp27}), the condition for the generalized Brillouin zone is given by
\begin{equation}
\left|\beta_1\right|=\left|\beta_2\right|.
\label{eqapp35}
\end{equation}
Now, by combining Eq.~(\ref{eqapp35}) and Vieta's formulas, we can get the absolute value of $\beta$ which gives the radius of the generalized Brillouin zone:
\begin{equation}
r^\prime=\sqrt{\left|\prod_{j=1}^N\frac{p\left(x_j\right)-\displaystyle\frac{i\delta}{4}\left(\lambda_1\left(x_j\right)+\lambda_2\left(x_{j+1}\right)\right)}{p\left(x_j\right)+\displaystyle\frac{i\delta}{4}\left(\lambda_1\left(x_{j+1}\right)+\lambda_2\left(x_j\right)\right)}\right|}.
\label{eqapp36}
\end{equation}
Finally, in the limit of $N\rightarrow\infty$, Eq.~(\ref{eqapp36}) can be calculated as
\begin{equation}
\lim_{N\rightarrow\infty}r^\prime=\exp\left(\frac{1}{2}\int_0^adx\Im\frac{\lambda_1\left(x\right)+\lambda_2\left(x\right)}{2p\left(x\right)}\right).
\label{eqapp37}
\end{equation}
This is the main result (see Eq.~(\ref{eq6})) of our study on non-Hermitian waves in the continuous system. Whether our results depend on other difference procedures or not is left as future work.

%
%

\section{\label{secD}Band gap in the toy model}
We show a way to derive Eq.~(\ref{eq10}). The toy model introduced is given by
\begin{equation}
p\left(x\right)=p,~\lambda_1\left(x\right)=\lambda_2\left(x\right)=i\lambda\sin^2\frac{2\pi}{a}x,
\label{eqapp41}
\end{equation}
and $v\left(x\right)=0$ in Eq.~(\ref{eqapp31}). In this case, from Eq.~(\ref{eqapp36}), the imaginary part of the complex Bloch wave number is calculated as
\begin{equation}
\Im\left(k\right)=-\frac{\lambda}{4p}.
\label{eqapp42}
\end{equation}
Now the Fourier coefficients of these functions are obtained as
\begin{equation}
P_n=p\delta_{0,n},~\Lambda_{1,n}=\Lambda_{2,n}=\frac{i\lambda}{2}\left(\delta_{0,n}-\frac{1}{2}\delta_{2,n}-\frac{1}{2}\delta_{-2,n}\right),
\label{eqapp43}
\end{equation}
and $V_n=0$, where $\delta_{n,m}$ is the Kronecker delta:
\begin{eqnarray}
\delta_{n,m}=\left\{ \begin{array}{ll}
0 & {\rm if}~n\neq m, \vspace{3pt}\\
1 & {\rm if}~n=m.
\end{array}\right.
\label{eqapp44}
\end{eqnarray}
Importantly, in order to calculate the gap $\Delta$ between the second band and the third band, we take only the states $\Psi\left(k\pm\frac{2\pi}{a}\right)$ as the basis of the secular equation (\ref{eq8}) because the other states do not contribute to the gap. Then we can get the eigenvalue equation
\begin{eqnarray}
\left| \begin{array}{cc}
\displaystyle C\left(k_+\right)+\frac{i\lambda}{2}k_+-\Xi & \displaystyle -\frac{i\lambda}{4}k                        \vspace{8pt}\\
\displaystyle -\frac{i\lambda}{4}k                        & \displaystyle C\left(k_-\right)+\frac{i\lambda}{2}k_--\Xi
\end{array}\right|=0, \nonumber\\
\label{eqapp45}
\end{eqnarray}
where $k_\pm=k\pm\frac{2\pi}{a}$, $C\left(k\right)\equiv pk^2$, and $\Xi\equiv\left(\frac{\omega}{c}\right)^2$. By solving Eq.~(\ref{eqapp45}) in the generalized Brillouin zone, we can get the second band $\Xi_2\left(\Re\left(k\right)\right)$ and the third band $\Xi_3\left(\Re\left(k\right)\right)$. Therefore, the gap between these bands at $\Re\left(k\right)=0$ can be given by
\begin{eqnarray}
\Delta&=&\Xi_3\left(0\right)-\Xi_2\left(0\right) \nonumber\\
&=&\frac{\lambda^2}{8p}.
\label{eqapp46}
\end{eqnarray}

%
%

\section{\label{secE}Transfer matrix}
We describe a way to get the generalized Brillouin zone (\ref{eq11}) in the photonic crystal by using a transfer matrix. First of all, we focus on the $n$th unit cell in the multilayer. In the dielectric medium with $\hat{\varepsilon}_1$, the eigenstate of the wave equation (\ref{eqapp17}) is expressed as a plane wave. Then the magnetic field in $\left(n-1\right)a<x<\left(n-1\right)a+d_1$ can be written as
\begin{equation}
H_0\left(x\right)=A_ne^{ik_{1,+}\left[x-\left(n-1\right)a\right]}+B_ne^{ik_{1,-}\left[x-\left(n-1\right)a\right]}.
\label{eqapp51}
\end{equation}
Similarly, in the dielectric medium with $\hat{\varepsilon}_2$, the magnetic field in $\left(n-1\right)a+d_1<x<na$ can be obtained by
\begin{equation}
H_0\left(x\right)=C_ne^{ik_{2,+}\left[x-\left(n-1\right)a-d_1\right]}+D_ne^{ik_{2,-}\left[x-\left(n-1\right)a-d_1\right]}.
\label{eqapp52}
\end{equation}
Here in Eqs.~(\ref{eqapp51}) and (\ref{eqapp52}), $k_{i,\pm}~(i=1,2)$ are wave numbers of plane waves in each dielectric medium for a given value of a frequency $\omega$, and they are given by
\begin{eqnarray}
&&k_{i,\pm}=\frac{k_y}{2\eta_{i,yy}}\left(\eta_{i,xy}+\eta_{i,yx}\right) \nonumber\\
&&\pm\frac{1}{2\eta_{i,yy}}\sqrt{k_y^2\left(\eta_{i,xy}+\eta_{i,yx}\right)^2-4\eta_{i,yy}\left[k_y^2\eta_{i,xx}-\left(\frac{\omega}{c}\right)^2\right]}. \nonumber\\
\label{eqapp53}
\end{eqnarray}
Furthermore, since the electric field is obtained from
\begin{equation}
{\bm E}\left({\bm r}\right)=-\frac{1}{i\omega\varepsilon_0}\frac{1}{\hat{\varepsilon}}{\bm\nabla}\times{\bm H}\left({\bm r}\right),
\label{eqapp54}
\end{equation}
in the photonic crystal, the $y$ component of the electric field can be expressed as
\begin{eqnarray}
&&E_y\left(x\right)=\frac{1}{\omega\varepsilon_0}\left\{A_nf_{1,+}e^{ik_{1,+}\left[x-\left(n-1\right)a\right]}\right. \nonumber\\
&&\left.+B_nf_{1,-}e^{ik_{1,-}\left[x-\left(n-1\right)a\right]}\right\}
\label{eqapp55}
\end{eqnarray}
in $\left(n-1\right)a<x<\left(n-1\right)a+d_1$ and
\begin{eqnarray}
&&E_y\left(x\right)=\frac{1}{\omega\varepsilon_0}\left\{C_nf_{2,+}e^{ik_{2,+}\left[x-\left(n-1\right)a-d_1\right]}\right. \nonumber\\
&&\left.+D_nf_{2,-}e^{ik_{2,-}\left[x-\left(n-1\right)a-d_1\right]}\right\}
\label{eqapp56}
\end{eqnarray}
in $\left(n-1\right)a+d_1<x<na$, where
\begin{equation}
f_{i,\pm}=-k_y\eta_{i,yx}+k_{i,\pm}\eta_{i,yy}~\left(i=1,2\right).
\label{eqapp57}
\end{equation}

Now, since both the magnetic field and the electric field are continuous at $x=\left(n-1\right)a+d_1$ and at $x=na$, we can obtain the conditions for the coefficients $a_n,b_n,c_n$, and $d_n$ as
\begin{eqnarray}
&&\left( \begin{array}{cc}
e^{ik_{1,+}d_1}        & e^{ik_{1,-}d_1}        \vspace{3pt} \\
f_{1,+}e^{ik_{1,+}d_1} & f_{1,-}e^{ik_{1,-}d_1}
\end{array}\right)\left( \begin{array}{c}
A_n \vspace{3pt}\\
B_n
\end{array}\right) \nonumber\\
&&=\left( \begin{array}{cc}
1       & 1       \vspace{3pt}\\
f_{2,+} & f_{2,-}
\end{array}\right)\left( \begin{array}{c}
C_n \vspace{3pt}\\
D_n
\end{array}\right)
\label{eqapp58}
\end{eqnarray}
and
\begin{eqnarray}
&&\left( \begin{array}{cc}
e^{ik_{2,+}d_2}        & e^{ik_{2,-}d_2}        \vspace{3pt} \\
f_{2,+}e^{ik_{2,+}d_2} & f_{2,-}e^{ik_{2,-}d_2}
\end{array}\right)\left( \begin{array}{c}
C_n \vspace{3pt}\\
D_n
\end{array}\right) \nonumber\\
&&=\left( \begin{array}{cc}
1       & 1       \vspace{3pt}\\
f_{1,+} & f_{1,-}
\end{array}\right)\left( \begin{array}{c}
A_{n+1} \vspace{3pt}\\
B_{n+1}
\end{array}\right).
\label{eqapp59}
\end{eqnarray}
Hence, from Eqs.~(\ref{eqapp58}) and (\ref{eqapp59}), the transfer matrix can be expressed as
\begin{eqnarray}
\left( \begin{array}{c}
A_{n+1} \vspace{3pt}\\
B_{n+1}
\end{array}\right)=T\left( \begin{array}{c}
A_n \vspace{3pt}\\
B_n
\end{array}\right).
\label{eqapp510}
\end{eqnarray}
Here since
\begin{eqnarray}
\left( \begin{array}{c}
A_{n+1} \vspace{3pt}\\
B_{n+1}
\end{array}\right)=\beta\left( \begin{array}{c}
A_n \vspace{3pt}\\
B_n
\end{array}\right)
\label{eqapp511}
\end{eqnarray}
for $\beta=e^{ik_xa}$ is established due to the spatial periodicity, $\beta$ is an eigenvalue of the transfer matrix $T$. This means that the absolute value of the eigenvalue of $T$ gives the radius of the generalized Brillouin zone $r$. We note that this is consistent with the result of Ref.~\cite{Kunst2019}. By combining the condition (\ref{eqapp34}) and
\begin{equation}
\det T=\beta_1\beta_2,
\label{eqapp512}
\end{equation}
we can get
\begin{equation}
r=\sqrt{\left|\det T\right|}.
\label{eqapp513}
\end{equation}
Finally, since we have
\begin{equation}
\det T=e^{i\left(k_{1,+}+k_{1,-}\right)d_1+i\left(k_{2,+}+k_{2,-}\right)d_2},
\label{eqapp514}
\end{equation}
the radius of the generalized Brillouin zone can be given as
\begin{equation}
r=\exp\left(\frac{k_y}{2}\Im\sum_{i=1}^2\frac{\varepsilon_{i,xy}+\varepsilon_{i,yx}}{\varepsilon_{i,xx}}d_i\right).
\label{eqapp515}
\end{equation}
Importantly, Eq.~(\ref{eqapp515}) matches Eq.~(\ref{eq11}).

%
%

\section{\label{secF}Zak phase}
For convenience, we write Eq.~(\ref{eq5}) as
\begin{equation}
\hat{\Theta}H\left(x\right)=\left(\frac{\omega}{c}\right)^2H\left(x\right).
\label{eqapp61}
\end{equation}
For Eq.~(\ref{eqapp61}), we define the inner product between two states $|\psi\rangle$ and $|\phi\rangle$ as
\begin{equation}
\langle\phi|\psi\rangle=\int_0^adx~\phi^\ast\left(x\right)\psi\left(x\right).
\label{eqapp62}
\end{equation}
In Eq.~(\ref{eqapp61}), from the Bloch theorem, the eigenstates can be expressed as
\begin{equation}
H\left(x\right)=e^{ik_xx}u_{k_x}\left(x\right),
\label{eqapp63}
\end{equation}
where $k_x$ is the complex Bloch wave number in the generalized Brillouin zone (\ref{eq11}). Then the periodic part of the Bloch function satisfies
\begin{equation}
\hat{\Theta}_{k_x}u_{k_x,n}\left(x\right)=\left(\frac{\omega_n}{c}\right)^2u_{k_x,n}\left(x\right),
\label{eqapp64}
\end{equation}
where $n$ is the band index. Furthermore $u_{k_x,n}\left(x\right)$ is normalized as
\begin{equation}
\langle u_{k_x,n}|u_{k_x,n}\rangle=1.
\label{eqapp65}
\end{equation}
In this case, we can define the Zak phase as
\begin{equation}
\theta_n=i\int_{0}^{2\pi}d\theta~\left\langle u_{k_x,n}\left(x\right)\left|\frac{d}{d\theta}\right|u_{k_x,n}\left(x\right)\right\rangle,
\label{eqapp66}
\end{equation}
where $\theta$ is the real part of $k_x$. When the system becomes Hermitian, it has the conventional Brillouin zone, and Eq.~(\ref{eqapp66}) is equivalent to the conventional Zak phase~\cite{Zak1989}. We note that the value of the Zak phase depends on the configuration of the unit cell. The Zak phase takes real values because of the normalization condition (\ref{eqapp65}). In addition, the Zak phase is defined in terms of modulo $2\pi$ under the gauge transformation $|u_{k_x,n}\rangle\rightarrow e^{i\gamma\left(k_x\right)}|u_{k_x,n}\rangle~(\gamma\left(k_x\right)\in{\mathbb R})$.

%
%

%
%

\end{document}